# Measurable & Scalable NFRs using Fuzzy Logic and Likert Scale

Nasir Mahmood Malik, Arif Mushtaq, Samina Khalid, Tehmina Khalil, Faisal Munir Malik

emailnasir@yahoo.com, coolafee@yahoo.com, noshi_mir@yahoo.com,
tehmina_khalil08@yahoo.com, faisy20@yahoo.com

*Department of Computer Science, Bahria University, Islamabad, Pakistan,*

info@bahria.edu.pk

*Abstract*—Most of the research related to Non Functional Requirements (NFRs) have presented NFRs frameworks by integrating non functional requirements with functional requirements while we proposed that measurement of NFRs is possible e.g. cost and performance and NFR like usability can be scaled. Our novel hybrid approach integrates three things rather than two i.e. Functional Requirements (FRs), Measurable NFRs (M-NFRs) and Scalable NFRs (S-NFRs). We have also found the use of Fuzzy Logic and Likert Scale effective for handling of discretely measurable as well as scalable NFRs as these techniques can provide a simple way to arrive at a discrete or scalable NFR in contrast to vague, ambiguous, imprecise, noisy or missing NFR. Our approach can act as baseline for new NFR and aspect oriented frameworks by using all types of UML diagrams.

## I. INTRODUCTION

The major objective of the paper is to discuss existing methodologies to identify, capture and specify NFRs and propose an improved approach. A novel hybrid approach is proposed for discretely measurable and scalable NFRs by applying Fuzzy Logic and Likert Scale. The proposed methodology will give comparison between vague, ambiguous, imprecise, noisy or missing NFRs to clear, precise and measurable NFRs. Also as requirements can be obtained either through open-ended or closed-ended ways, or a combination of the two, the use of Likert scale can be very useful for gaining quantifiable requirements. It is also a regular method followed by Project Management Institute (PMI) for capturing various grey areas of requirements. The paper not only addresses yes/no questions of requirements but also focuses on areas between FRs and NFRs by using fuzzy logic and Likert Scale. Our work initially investigates work done related to NFRs along with background of issues. Secondly we have given our proposed approach, findings and finally expected future work as well.

## II. RELATED WORK:

Research related to NFRs frameworks in UML mostly used use case and class diagrams and almost all of them have proposed extensions in the current UML for the incorporation of NFRs. Moreover, past research works do have serious issues at the point of integration of FRs and NFRs, which are mainly due to the separation of concerns for both types of requirements. How FRs and NFRs should be developed in a tightly or loosely integrated approach is another area of future research. As at some point of time you might need them separately and yet integrated to address the overall objectives of the system.

In reality, there is no formal rule to form/analyze aspects and to address cross cutting concerns for FRs and NFRs. Integrating NFRs with FRs at requirement level through some integration point is a normal practice as we found from our survey [Table-1]. Also as per our literature survey many authors have used use case diagrams for this integration and most of them have proposed extensions to existing UML model for the integration purpose [Table-1]. Furthermore, these approaches have kept the options open both for functional and non functional view of the system.

However, NFR-Based approach also emphasizes the use case cohesion. Aspect-Based approaches concentrate more on class diagrams. Our approach includes the proposal of measurement of NFRs i.e. possibility of measurement of NFRs is explored through Measurable NFRs and Scalable NFRS and we have also tried to reduce the grey areas of NFRs.

## III. PROPOSED APPROACH

The research related to Non Functional Requirements (NFRs) emphasize its importance but very little implementation could be found. By definition, requirement should be something which can be verified at some point of time, which is not possible for all the non functional requirements i.e. you can not formally say yes or no e.g. usability. Also there is no point to be precise when we don't know what is actually required and what NFRs are necessary for which system. So, our research is aimed at bridging this gap between vague/hazy and precise/scalable. To make non functional requirements measurable we have used fuzzy logic and Likert Scale.

TABLE-1: RELATED WORK

| Research Work | UML Diagrams Used | | | | | Approach Used |
|---|---|---|---|---|---|---|
| | Class | Use case | Sequence | State | Extension Needed | |
| **NFR Framework [01]** | | | | | | NFR Based |
| **Model for early quality attribute [02]** | | √ | √ | √ | √ | NFR Based |
| **Composition pattern in design phase [03]** | | | √ | | √ | Aspect Based |
| **UML-Based Performance Engineering [04]** | | √ | | √ | √ | NFR Based |
| **AORE Model [05]** | | √ | | | √ | Aspect Based |
| **Extending UML with UML profile [06]** | | √ | | | √ | NFR Based |
| **UMLAUT Framework[07]** | √ | | | | √ | Aspect Based |
| **A use-case and goal-driven approach [08]** | | √ | | | √ | NFR Based |
| **NFRs in software architecture [09]** | N/A | N/A | N/A | N/A | N/A | NFR Based |
| **Aspect support in design phase [10]** | √ | | | | | Aspect Based |

*A. NFR using Likert Scale:*

The Likert Scale was developed in 1932 by Rensis Likert [11]. These scales always ask to indicate how much they agree or disagree, approve or disapprove, believe to be true or false. There isn't any perfect approach to follow the Likert Scale, however the mot important thing to have a balanced overview of the situation, at least 5 response categories may be included. A Likert Scale adds up responses to statements representative of a particular attitude. Likert Scale allows a participant to provide feedback that is slightly more expansive than a simple close-ended question but that is much easier to quantify than a completely open-ended response. The numerous advantages of a Likert scale are obvious i.e. they are 'easy' to construct, administer and score. Likert Scale lists a set of statements and provides a 5-point, 6-point, 7-point scale and gives each cell a value.

Another issue with Non-Functional Requirements (NFRs) is that most of the times they are stated in natural language. As by definition, each requirement, functional or non functional, must be objective and quantifiable. Secondly there must be some way to measure whether the requirement has been met or not. According to [2], NFR frameworks can be based on Goals. The research begins with identification of hard and soft goals that represent NFR which stakeholders agree upon. Hard goals are easy to incorporate (becomes functional requirements).

Whereas Soft goals are goals that are hard to express, but tend to be global qualities of a software system. These could be reliability, usability, performance, security, maintainability, flexibility etc (mostly non functional requirements) in a given system. These soft goals are then usually decomposed into sub goals making tree type structures. Most of the times they are conflicting and one way to reach on some measurable scale is the use of Likert Scale until all the root soft goals are satisfied. The detail regarding the approach is presented in [2].

TABLE-2: LIKERT SCALE EXAMPLE:

| Sr. # | Available Options | Assigned Value |
|---|---|---|
| 1 | Strongly Agree | 7 |
| 2 | Agree | 6 |
| 3 | Agree Somewhat | 5 |
| 4 | Undecided | 4 |
| 5 | Disagree Somewhat | 3 |
| 6 | Disagree | 2 |
| 7 | Strongly Disagree | 1 |

The proposed methodology will help in quantifying requirements from simple yes/no to reach more comprehensive feedback. For example, **I want to close this project in one month?** We can proceed in an open ended way and get some descriptive response but better line of action could be the Likert Scale approach. The options according to Likert Scale are shown in Table-2. In response to the aforesaid question, a list of say 30 possible answers to the topic would be made up. For instance, in response to our question we receive 15 favorable, and 15 unfavorable answers. Participants would then rate each statement on a seven-point scale as follows: A person's attitude is the summed score from each question.

A Likert Scale will give each cell a value either in ascending or descending order. In the above Likert Scale example Strongly Agree has a value of 7, Agree, 6, Agree somewhat 5, Undecided 4, Disagree somewhat 3, Disagree 2 and Strongly disagree 1. We may sum up, value associated with each of our 30 respondent's answers to a particular question. In this case a high score would indicate a favorable attitude, towards the closure of the project and a low score an unfavorable attitude towards the closure of the project. So this way reaching a concrete decision becomes much easier through quantifiable what to do with the fate of the project rather we have reached our decision from non measurable and vague to precise and predictable decision regarding the decision of the project.

*B. NFR using Fuzzy Logic:*

A formal approach and a practical method are developed to analyze the complex relationships between requirements. Conflicting requirements can be identified and represented using both qualitative terms and quantitative measures. Multiple requirements with complex relationships among them are fused into an overall system requirement based on fuzzy multi-criteria decision techniques [12]. The approach has two major challenges with requirement analysis in concurrent engineering are: (1) requirements from multiple members of a concurrent engineering team are often conflicting with each other; and (2) requirements are often imprecise in nature. Existing formal methods for requirement engineering are very limited in addressing these issues.

This paper is a step in this direction to reach on precise requirements if not 100%. Also [13] proposed approach provides a framework for formally analyzing and modeling conflicts between requirements, and for users to better understand relationships among their requirements on fuzzy techniques. Initially Fuzzy Logic (FL) was conceived as a better method for sorting and handling data but it has proven to be an excellent choice for many control system applications due to human control logic. NFRs require all it takes to define precise and scalable non functional requirements.

The approach has been followed by professional organizations such as Project Management Institute (PMI) has made Likert Scale as standard to measure requirements present in grey areas. Our research work proposes that both of these approaches will be very effective for obtaining consistent and complete non functional requirements. In order to lessen the noise and ambiguity in NFRs, fuzzy logic could be helpful.

The major issue of NFRs has been the quantization errors i.e. measurement of NFRs. Especially measurement between the lines. Our methodology can measure NFRs e.g. performance measurement for general database transactions taking place may require several ranges to be defined for example "slow", "average" and "fast". Each range could then be a function of number of transactions. The truth value could then be decided as "somewhat slow", "slightly average" or "not fast" or the scale as shown in Figure-1.

This will give clear and more precise understanding of requirements, rather this will no focus simply on yes/no answer or Boolean answer it will give variations with answers. According to Figure-1 some users might be satisfied with only 70% of questions while others with 30% only i.e. more quantization without error. The paper emphasizes requirements phase only, however latest publications also include approaches as proposed in [14] for optimizing designs based on various NFRs.

*C. Evaluation of Reliability NFR*

During the process of gathering data related to common government applications it transpired that reliability NFR can be divided into two Sub-NFRs (frequency of failure and recoverability). The results obtained from this survey give us following wait, i.e. for recoverability 8 and 2 for frequency of failure. Now, if the likelihood of software to fail is at the level of say 0.1 or less, then it is acceptable as a highly reliable software. Likewise, probability of recoverability is say 0.8 or more, the software will be deemed to be highly reliable. On similar lines, reliability could be measured using a template given in Table-2 and the ideal score will be 8.

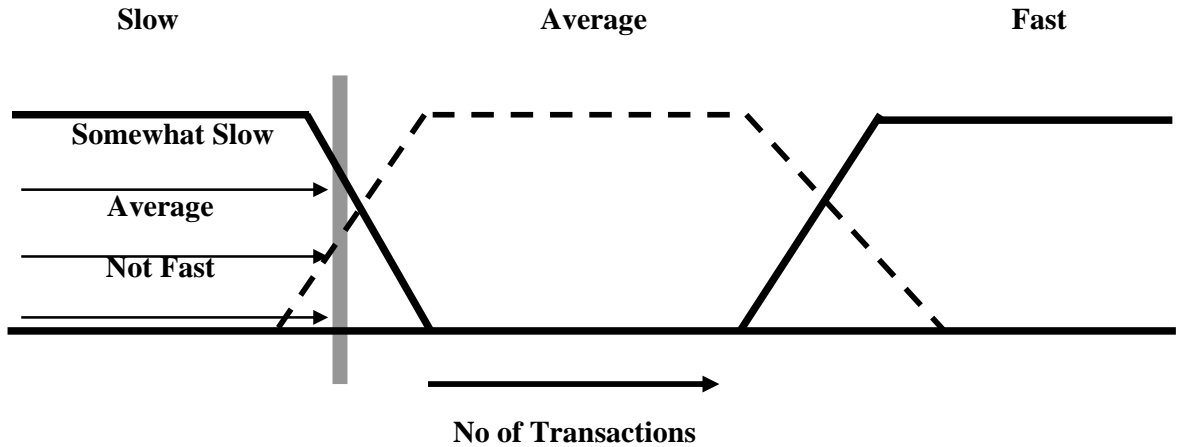

Figure-1: Handling of Vague Requirements: Fuzzy Logic Example

TABLE-3: FUZZY LOGIC EXAMPLE

| Reliability Status | Frequency of Failure | Recoverability |
|---|---|---|
| High Reliable | 0.1 or Less (Low) | 0.8 or More (High) |
| Reliable | Between 0.1 and 0.2 (Average) | Between 0.7 and 0.8 (Average) |
| Not Reliable | More than 0.2 (High) | Less than 0.7 (Low) |
| **Weightage** | 2 | 8 |

However, depending upon requirements, any organization can mold this template for three things i.e. highly reliable, average and low reliable. From the table it is obvious that more emphasis is on recoverability that is why it has been assigned the highest weightage 8. But in an ideal scenario the frequency of failure should be as low as possible (lowest will be zero). Therefore, recoverability should be as high as possible (max could be 10).

IV. FINDINGS

According to Somerville [2], measurement of products or systems is absolutely fundamental to the engineering process. We are convinced that measurement as practiced in other engineering disciplines is *IMPOSSIBLE* for software engineering. This statement clearly highlights the issues related with all requirements associated with software system. Therefore, our idea behind the research is to make NFRs measurable. Moreover, if we could measure some non functional requirements in absolute terms e.g. response time should be fast; is a vague NFR, whereas, if we say response time should be less than one second, it will become a measurable non functional requirement. So resultantly it can be treated like a normal functional requirement. So, basically our research is going to divide integration point of FRs and NFRs into three requirements. The approach is explained in Figure-2 i.e.

a. Functional Requirements (FRs)
b. Measurable NFRs (M-NFRs)
c. Scalable NFRs (M-NFRs)

Figure-2: Proposed Approach

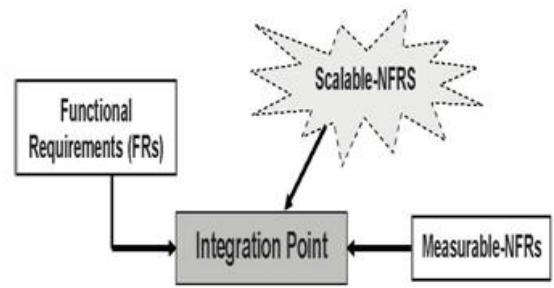

Integration of functional and non functional requirements is still a major area of research. The main reason for both types of requirements to be treated separately is their separate area of concern. Comparison of various approaches (Table-1) shows that both FRs and NFRs should be treated separately but do have some integration point, through which we can have two distinct views of the system. FRs deal with the actually functionality of the system whereas NFRs deal with the constraints that are affecting this functionality. The methodology can help to achieve more realistic requirements that is acceptable to the customers and feasible to implement for developers by fully exploring the grey areas of NFRs. Also, explicit specification of imprecise requirements provides a basis for verification and validation of software systems.

## V. CONCLUSION

Non functional requirements are considered as contradictory & vague, difficult to implement & enforce during development and validate/verify/evaluate for the customer prior to delivery of the actual system. So this paper is a step in this direction to get precise and clear non functional requirements (NFRs) along with functional requirements (FRs) at an early stage. The existing research on integration of FRs and NFRs mainly concentrates on using UML diagrams with proposed extensions for capturing NFRs. In contrast the methodology being proposed here is capturing discretely measurable or scalable NFR, in contrast to vague, ambiguous, imprecise, noisy or missing NFR based on fuzzy logic and Likert scale. Furthermore this proposed methodology could act as a baseline for new NFR and aspect based frameworks by using all types of UML diagrams to get precise and accurate NFRs.

**Future Work**

The future work will include incorporation of this approach into practical software applications i.e. case studies will be processed and evaluated. The detailed evaluation will be obtained through the results of these case studies and comparison of this approach with other similar approaches.